\documentclass[fleqn,twoside]{article}
\usepackage{espcrc2}

\usepackage{epsfig}

\newcommand{\AmS}{{\protect\the\textfont2
  A\kern-.1667em\lower.5ex\hbox{M}\kern-.125emS}}

\title{High Energy Description of Processes with Multiple Hard Jets}

\author{Jeppe R.~Andersen\address{Theory Division, Physics
   Department, CERN, CH-1211 Geneva 23, Switzerland}
   and
   Jennifer M.~Smillie\address{Department of Physics, UCL, 
     Gower Street, WC1E 6BT, United Kingdom}.
}

\begin{document}

\begin{abstract}
  \emph{High Energy Jets} (HEJ) is a new framework for approximating the
  all-order perturbative corrections to multi-jet processes, with a focus on
  the hard, wide-angle QCD emissions, which underpins the perturbative
  description of hard jets. In this contribution we review the basic concepts
  of \emph{HEJ}, and present some new predictions for observables in
  dijet-production, and for W-boson production in association with at least 3
  jets.
\vspace{1pc}
\end{abstract}

\maketitle

\section{INTRODUCTION}
Since the production cross-section at the LHC for particles charged under QCD
generally will be larger than that for colourless particles, many of the
discovery channels used in the search for new physics involves the detection
of hard, hadronic jets. The large mass hierarchy between any (necessarily
heavy, in order to avoid existing exclusion limits) produced new particle and
those of the decay products means often many such jets should be produced in
the decay of a new state. The finger prints of any such new physics will,
however, have to be found amongst a large contribution to the same signal from
multi-jet processes within the Standard Model. Therefore, a detailed
understanding of the Standard Model processes will assist in the search for
new physics. Examples of Standard Model processes acting as background to
many searches are e.g.~$W,Z+$jets (especially with 3,4 jets or more).

However, even the nature of some Standard Model processes is best studied in
events with jets. For example, the $CP$-structure of the induced Higgs boson
couplings to gluons through a top-loop could be measured by a study of the
azimuthal angle between the two jets in events with a Higgs boson in
association with dijets\cite{Klamke:2007cu,Andersen:2010zx}.

In both examples, hard radiative corrections will be sizeable at the LHC, by
which we mean that the exclusive $(n+1)$-jet rate is a significant component
of the inclusive $n$-jet rate. Therefore, a tree-level description of the
$n$-jet rate will be insufficient for a satisfactory description of the final
state (even if dressed with a parton shower). The reason for the increased
importance in many situations of hard, perturbative corrections at the LHC
over the situation at previously, lower energy colliders is very simple. Two
effects act to suppress hard corrections: the increasing powers of the
perturbative coupling, and the necessary increase in the light-cone momentum
fraction of the partons extracted from the proton beyond that necessary for
the final state without the additional hard jet. The suppression from this
last kinematic effect is caused by the decrease in the parton density
functions (pdf) as the light-cone momentum fraction $x$ is
increased. However, for processes with at least two particles in the final
state, there is a fine trade-off between the suppression from the pdf and the
phase space for additional emission (even when this is hard in transverse
momentum), as the rapidity span between the two particles is increased. At
previous, lower-energy colliders, this balance was tipped more towards a
suppression than will be the case at the LHC. At previous colliders, the ``significant''
rapidity separation of the two objects which is necessary for the opening of
phase space for additional radiation would already bring the light-cone
momentum fractions into the region of extremely fast falling pdfs as $x\to1$,
thus effectively vetoing additional emissions. However, the situation is
different for the LHC processes discussed above, since a significant rapidity
separation will be directly imposed in the analysis of the $CP$-properties of
the Higgs boson couplings, and in the case of $W$-boson production with at
least 3 jets, two jets will naturally be produced with a size-able separation
in rapidity\cite{Binoth:2010ra}.

The considerations above are just simple kinematical observations, which hold
true for any multi-particle process, and any reasonable theoretical
description thereof. However, the amount of hard emissions in the span
between the two extremal (in rapidity) particles will differ between:-
\begin{itemize}
\item[]\textbf{Processes:} according to whether there is a colour octet
  channel between the two particles (i.e.~the possibility of a gluon
  exchange)\cite{Dokshitzer:1991he}. This induces a difference in the
  radiation pattern between e.g.~the weak-boson- and gluon-fusion channel in
  $hjj$.
\item[]\textbf{Theoretical models:} A \emph{fixed order calculation} like NLO
  will obviously be able to generate just one hard jet beyond that of the LO
  process, irrespectively of the length of the rapidity span between the two
  extremal jets and the connected growth of the phase space for additional
  emissions.

  A \emph{parton shower} description will generally underestimate the amount
  of hard radiation (and thus the number of jets) in-between the two extremal
  jets, although the description can be improved by a few fixed orders
  through a CKKW-style
  matching\cite{Catani:2001cc,Mangano:2001xp,Lonnblad:1992tz}, or to full NLO
  accuracy\cite{Frixione:2002ik,Alioli:2010xd}, so far for processes of low
  multiplicity only.
\end{itemize}

\begin{figure}[bth]
  \centering
  \epsfig{width=\columnwidth,file=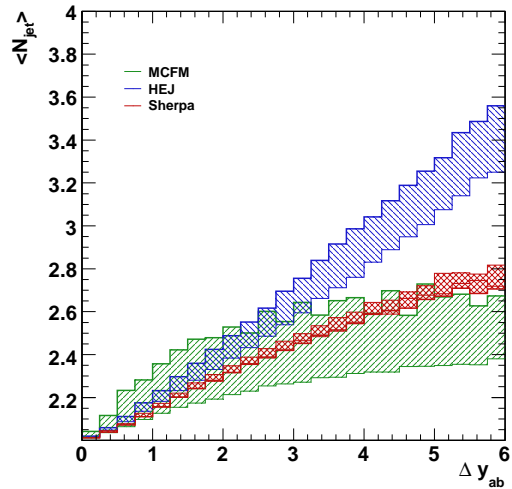}
  \caption{The average number of hard jets ($p_\perp>40$~GeV) in events with
    a Higgs boson and at least two jets at a 10~TeV $pp$-machine, as a
    function of the rapidity difference $\Delta y_{ab}$ between the most
    forward and most backward hard jet. The predictions from three different
    models are compared: Fixed order NLO by MCFM\cite{Campbell:2010cz}
    (green), the SHERPA shower Monte Carlo\cite{Gleisberg:2008ta}, including
    CKKW-matching with processes for a Higgs-boson and up to 4 final state
    partons (red), and finally the all-order framework of \emph{High Energy
      Jets}\cite{Andersen:2008ue,Andersen:2008gc,Andersen:2009nu,Andersen:2009he}
    (blue).  See Ref.\cite{Binoth:2010ra} for more details.}
  \label{fig:avgjetshjj}
\end{figure}

In Fig.~\ref{fig:avgjetshjj} we show the predictions obtained in three
different approaches for the average number of hard jets ($p_\perp>40$~GeV)
in events with a Higgs boson and at least two jets at a 10~TeV $pp$-machine,
as a function of the rapidity difference $\Delta y_{ab}$ between the most
forward and most backward hard jet. The bands describe the uncertainty of
each prediction as estimated by varying the renormalisation and factorisation
scales by a factor of two. The scale choice is identical in the calculations
of MCFM\cite{Campbell:2010cz} and High Energy
Jets\cite{Andersen:2008ue,Andersen:2008gc,Andersen:2009nu,Andersen:2009he}
(described in the next section), whereas the SHERPA\cite{Gleisberg:2008ta}
setup chooses scales according to the CKKW-procedure (with a much smaller
range induced by varying the scale by a factor of two). See
Ref.\cite{Binoth:2010ra} for more details on the comparison. All models
obviously agree on the increasing importance of hard radiative corrections
with increasing rapidity span\footnote{We note in passing that a large
  rapidity span (3-4 units) between two jets is often \emph{required} in the
  analysis of $hjj$, also in analysis of the gluon fusion (GF)
  channel\cite{Klamke:2007cu}.}. At smaller rapidity spans ($\Delta
y_{ab}<2$), where the phase space for many hard emissions is suppressed, the
predictions agree between each other. However, as the rapidity span is
increased (into the region of interest of at least 3-4 units), differences
start to emerge, since the models which can reach higher multiplicities start
becoming sensitive to these. An obvious observation is that with an average
number of jets of around 2.5 in the region of interest, the exclusive two and
three jet rates in the NLO calculation are equally large. It would seem
mandatory to care about the description of further hard radiation, if one is
concerned with any observable which depends on the configuration of the final
state. Even the ``total cross section'' can depend crucially on the jet
configuration after acceptance
cuts have been imposed (see Ref.\cite{Binoth:2010ra} for a discussion on the
estimated effect of jet vetos as obtained in the three models).

A reliable understanding and description of the final state in terms of jets
is obviously desirable, in order to e.g.~form observables for the extraction
of the $CP$-properties in $hjj$ which are stable against the higher order
corrections\cite{Andersen:2010zx}, or to help in further discriminating
Standard Model contributions to the search channels for new physics, by
isolating the regions in phase space where the amount of hard radiative
corrections occur for the SM process (like illustrated above in the example
of GF contribution to $hjj$).

\section{HIGH ENERGY JETS (HEJ)}
\label{sec:high-energy-jets}
The all-order perturbative framework of \emph{High Energy Jets} (\emph{HEJ})
developed in
Ref.\cite{Andersen:2008ue,Andersen:2008gc,Andersen:2009nu,Andersen:2009he} is
addressing the short-comings in the description of multiple hard,
perturbative corrections in both the (low) fixed-order and in the parton
shower formulation. The perturbative description obtained with \emph{HEJ}
reproduces the correct, all-order, full QCD, limit for both real and virtual
corrections to the hard perturbative matrix element for the hard, wide-angle
emissions which underpins the perturbative description of the formation of
additional jets. The central parts of the formalism were presented in
Ref.\cite{Andersen:2009nu,Andersen:2009he} and discussed further in
Ref.\cite{Andersen:2010ch}. In the following, we will first give just a brief
overview of the underlying formalism. Since two comparative studies of
results obtained with (CKKW-matched) shower, NLO and HEJ for $hjj$ and $Wjj$
were already reported in Ref.\cite{Binoth:2010ra}, we will here discuss new
results obtained from the application of HEJ to the LHC processes of dijets
and $W$+3 jets. These will form parts of forthcoming publications.

\boldmath
\subsection{Dominance of the $t$-channel poles}
\label{sec:dominance-t-channel}
\unboldmath 

The limit of pure $N$-jet amplitudes for large invariant mass between each
jet of similar transverse momentum is described by the
FKL-amplitudes\cite{Fadin:1975cb,Kuraev:1976ge}, which are at the foundation
of the BFKL framework\cite{Balitsky:1978ic}. The physical picture arising
from the FKL amplitudes is one of effective vertices connected by $t$-channel
propagators. The reduction of the formalism to the two-dimensional BFKL
integral equation relies on many kinematical approximations, which are
extended to all of phase space. Using an explicit (or so-called iterative)
solution to the BFKL equation\cite{Andersen:2001kt}, it is however
straightforward to show that despite the logarithmic accuracy, the
perturbative expansion of the (B)FKL solution does not give a satisfactory
description of the results obtained order by order with the true perturbative
series from QCD\cite{Andersen:2008gc}.

\emph{High Energy Jets}\cite{Andersen:2009nu,Andersen:2009he} inherits the
idea of effective vertices connected by $t$-channel currents in order to
reproduce the correct limit of $N$-jet amplitudes, but goes beyond
controlling just the logarithmic accuracy of the FKL formalism. The
kinematic building blocks of the FKL formalism depend on transverse momenta
only, as a result of the kinematic limits applied in order to separate the
amplitude into effective vertices separated by $t$-channel
exchanges\cite{Fadin:2006bj}. We will discuss how to obtain such separation
without resorting to kinematic approximations.

The $2\to2$ scattering $qQ\to qQ$ obviously proceeds through just a
$t$-channel exchange of the gluon current generated by a quark. A careful
analysis\cite{Andersen:2009nu,Andersen:2009he} of the helicity structure in
$qg\to qg$ and $gg\to gg$-scattering reveals that all the amplitudes, where
the helicity of the gluon is unchanged\footnote{All helicity-flip amplitudes
  are systematically suppressed by a factor of $\hat s$.}, factorise again
into two currents, contracted over a $t$-channel pole. This allows for a
definition of the current exchanged in the $t$-channel also for scatterings
involving gluons. The emission
of additional gluons is performed by gauge-invariant\footnote{by which we of
  course mean fully gauge invariant, not just up to sub-asymptotic terms as
  it is often meant in the BFKL literature.}, effective vertices. The virtual
corrections are approximated with the \emph{Lipatov ansatz} for the
$t$-channel gluon propagators (see Ref.\cite{Andersen:2009nu} for more
details). The end result is a formalism which provides a good approximation
order-by-order to the full QCD results, while being sufficiently fast to
evaluate that all-order results for the amplitudes can be explicitly
constructed and integrated over the $n$-body phase spaces (with an upper
limit on $n$ sufficiently high to guarantee convergence).

\subsection{Matching to Fixed Order}
\label{sec:matching-fixed-order}
In the cases of low jet multiplicity (up to 4), where the exact tree-level
amplitudes are known, the formalism is matched to this accuracy by mapping
the generated $n$-jet, $m$-parton configuration into a configuration of $n$
on-shell final-state partons, for which the tree-level amplitudes can be
evaluated. The all-order event weight is then multiplied by the ratio of the
full and the approximate tree-level amplitude. The low-multiplicity,
tree-level amplitudes are evaluated using MadGraph\cite{Alwall:2007st}.

\section{RESULTS}
\label{sec:results}
In the following, we will presents results obtained for processes of pure
jets (inclusive dijets) and for W-production in association with at least 3
jets. We will see that the increasing relevance of hard, perturbative
corrections with increasing rapidity span (as indicated in
Fig.~\ref{fig:avgjetshjj}) is completely general for all the processes under
consideration. The exact rate of the increase depends on the jet cuts and
definition, and the pdfs (i.e.~whether it is a processes dominated by a $gg$
(as in the case of dijets) or $qg$ ($W$+jets) initial state), with
$gg$-dominated processes being less dominated by hard, radiative corrections
because of the steeply falling gluon pdf. In the case of pure jets, we will
discuss how the effect of the evolution of the amount of real radiation with
the rapidity span can be measured directly with the data of the first year of
running with the LHC by measuring $\mathrm{d}\sigma/\mathrm{d}\phi_{fb}$
(where $\phi_{fb}$ is the azimuthal angle between the forward/backward hard
jet) in bins of the rapidity difference between the forward/backward hard
jet.

While an increasing rapidity span clearly forces the increasing relevance of
hard, perturbative corrections, the importance of resumming such corrections
to get a stable, perturbative description of the final state is obviously not
limited to the study of increasing rapidity spans. In the case of $W+$3 jets
we will see that the tail of the $H_T$-distribution also attracts large
contributions from hard, perturbative corrections.

\subsection{Pure Jets}
\label{sec:pure-jets}
\begin{figure}[bt!]
  \centering
  \epsfig{width=\columnwidth,file=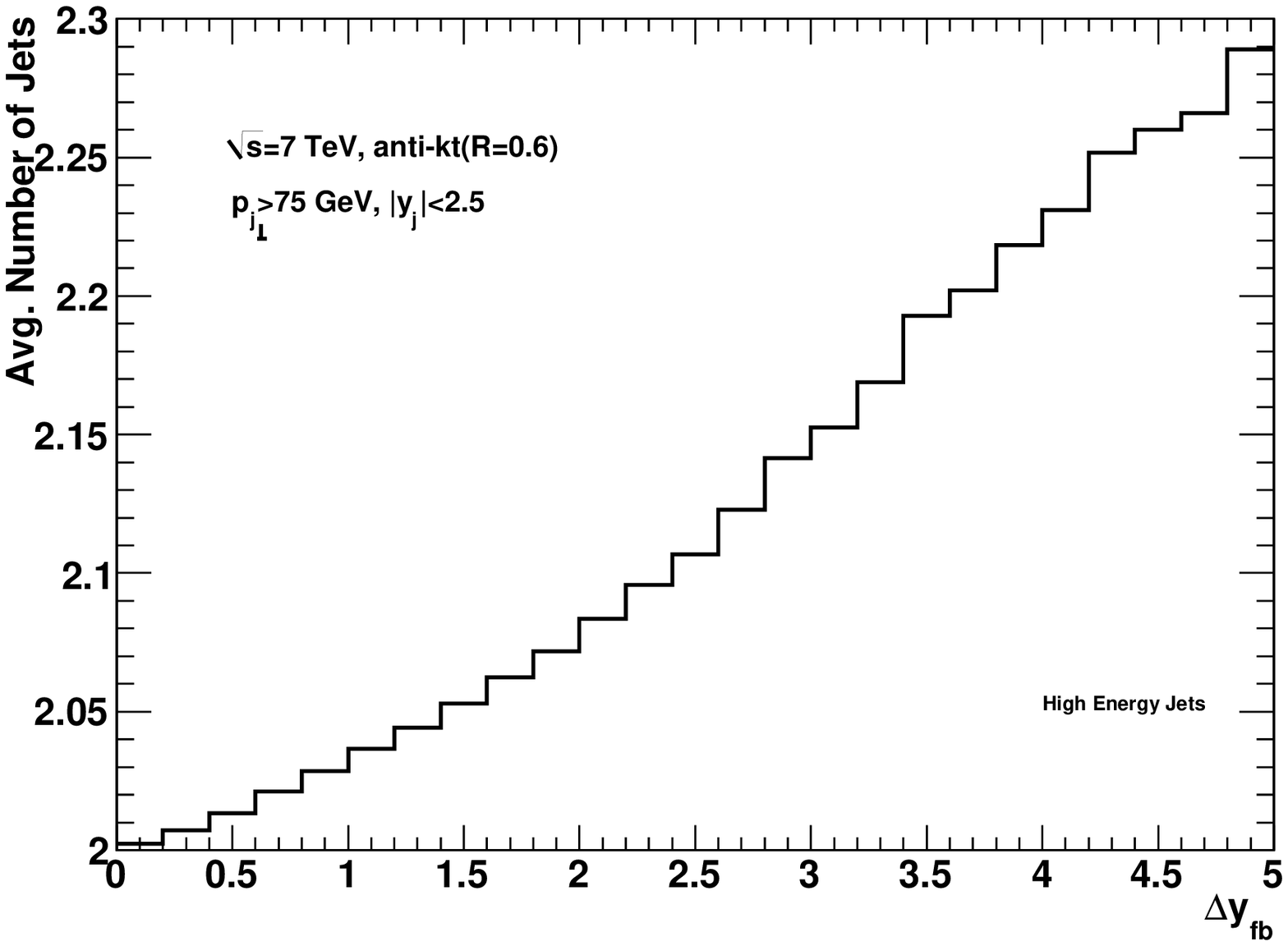}
  \epsfig{width=\columnwidth,file=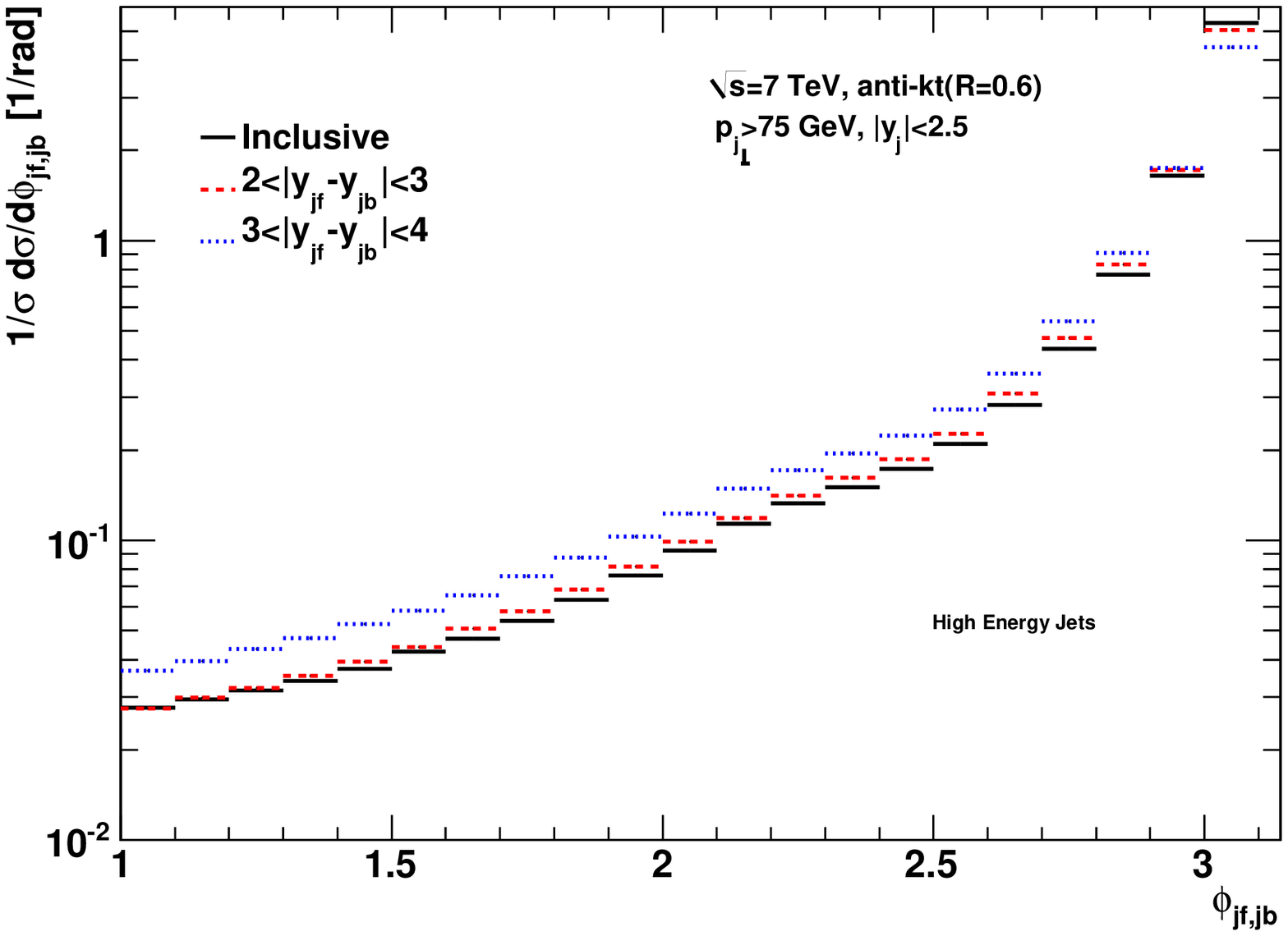}
  \caption{Top: The average number of hard jets ($p_\perp>75$GeV) in dijet
    events at the LHC (7~TeV). Bottom: The increasing weight of hard,
    radiative corrections for increasing rapidity spans between the
    forward/backward hard jet leads to the distribution on the azimuthal
    angle between these jets to be less peaked at the back-to-back
    configuration.}
  \label{fig:jjdsdphi}
\end{figure}

In this study, we present results for dijet-production at a 7~TeV
pp-collider. The jet-algorithm is anti-kt, with an $R$-parameter of 0.6, and
the transverse momentum of the jets are required to be harder than 75GeV,
with an absolute rapidity less than 2.5. Earlier analyses of results obtained
using the less accurate BFKL formalism\cite{Andersen:2003gs} already
indicated there should be a strong dependence between the average number of
jets and the rapidity span between the most forward/backward hard jet. In
Fig.~\ref{fig:jjdsdphi} (top) we show the prediction obtained using
\emph{HEJ} for the correlation between the average number of hard (all above
75GeV in transverse momentum) jets and the rapidity span between the most
forward/backward hard jet.

Clearly, the increasing importance of hard, radiative corrections will impact
many observables and event shapes. In Fig.~\ref{fig:jjdsdphi} (bottom) we
show the results for $1/\sigma\ \mathrm{d}\sigma/\mathrm{d}\phi_{fb}$ (where
$\phi_{fb}$ is the azimuthal angle between the forward/backward hard jet) for
three bins of the rapidity difference $y_{fb}$ between the forward/backward
jet. For increasing rapidity spans, the increasing amount of hard radiative
corrections leads to a distribution which is less peaked at the situation of
back-to-back jets.

\subsection{W+3 Jets}
\label{sec:w+jets}
The formalism of \emph{HEJ} supersedes the less accurate BFKL description of
W+jets implemented in Ref.\cite{Andersen:2001ja}: not only does \emph{HEJ}
include matching to fixed order results, but the matching corrections are
much smaller than those which would arise in a pure BFKL approach.

\begin{figure}[htbp!]
  \centering
  \epsfig{width=\columnwidth,file=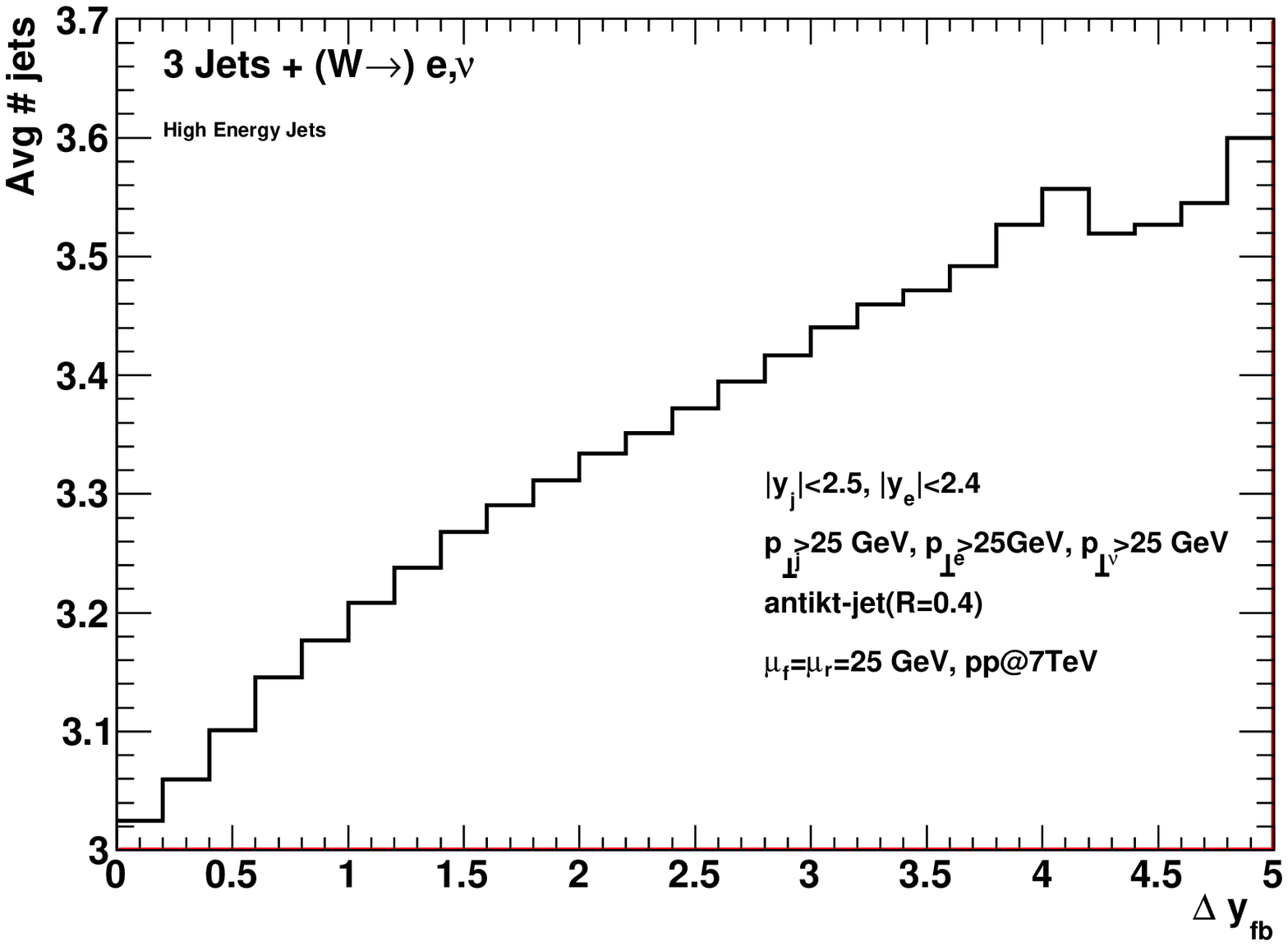}
  \epsfig{width=\columnwidth,file=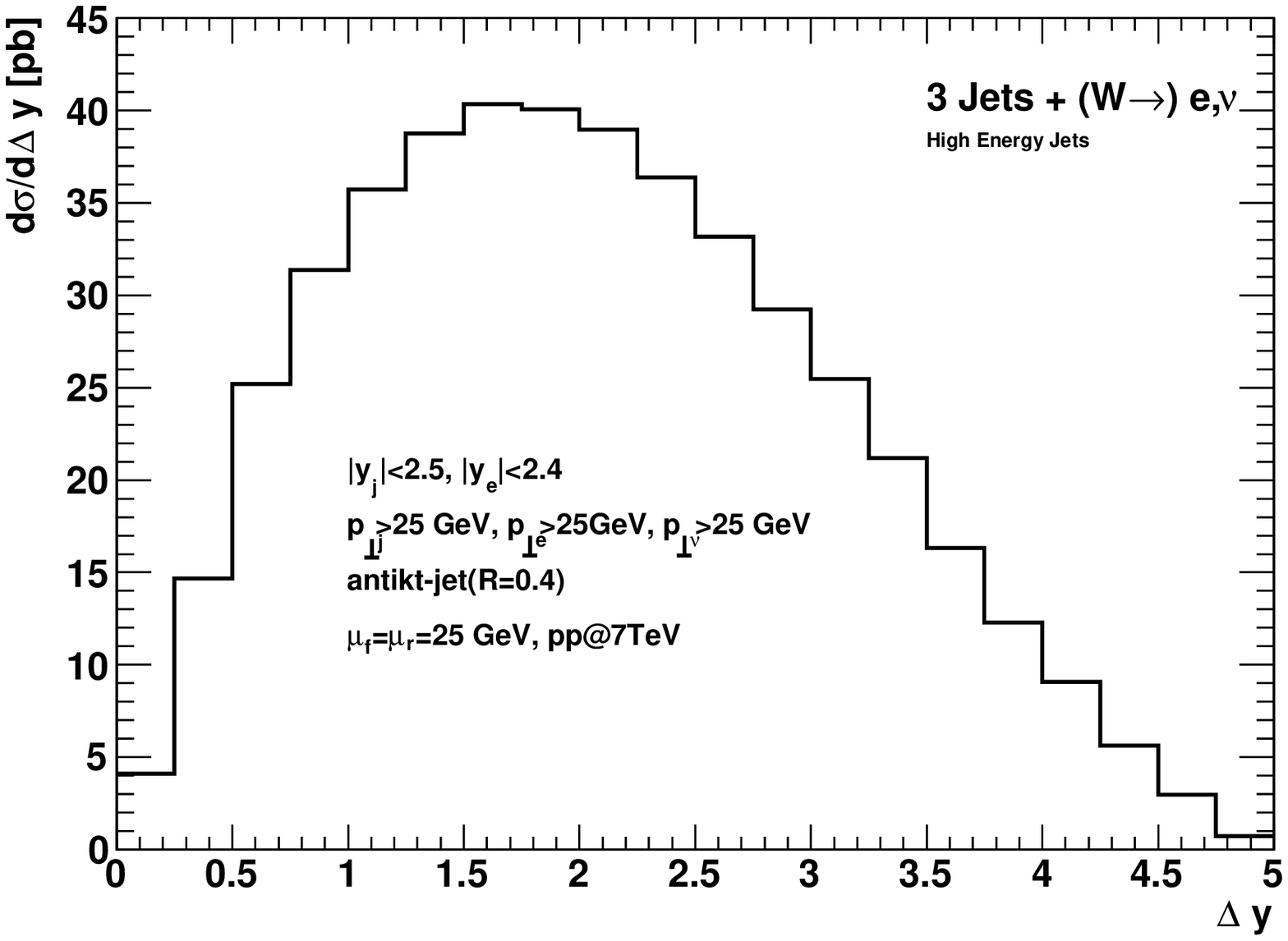}
  \epsfig{width=\columnwidth,file=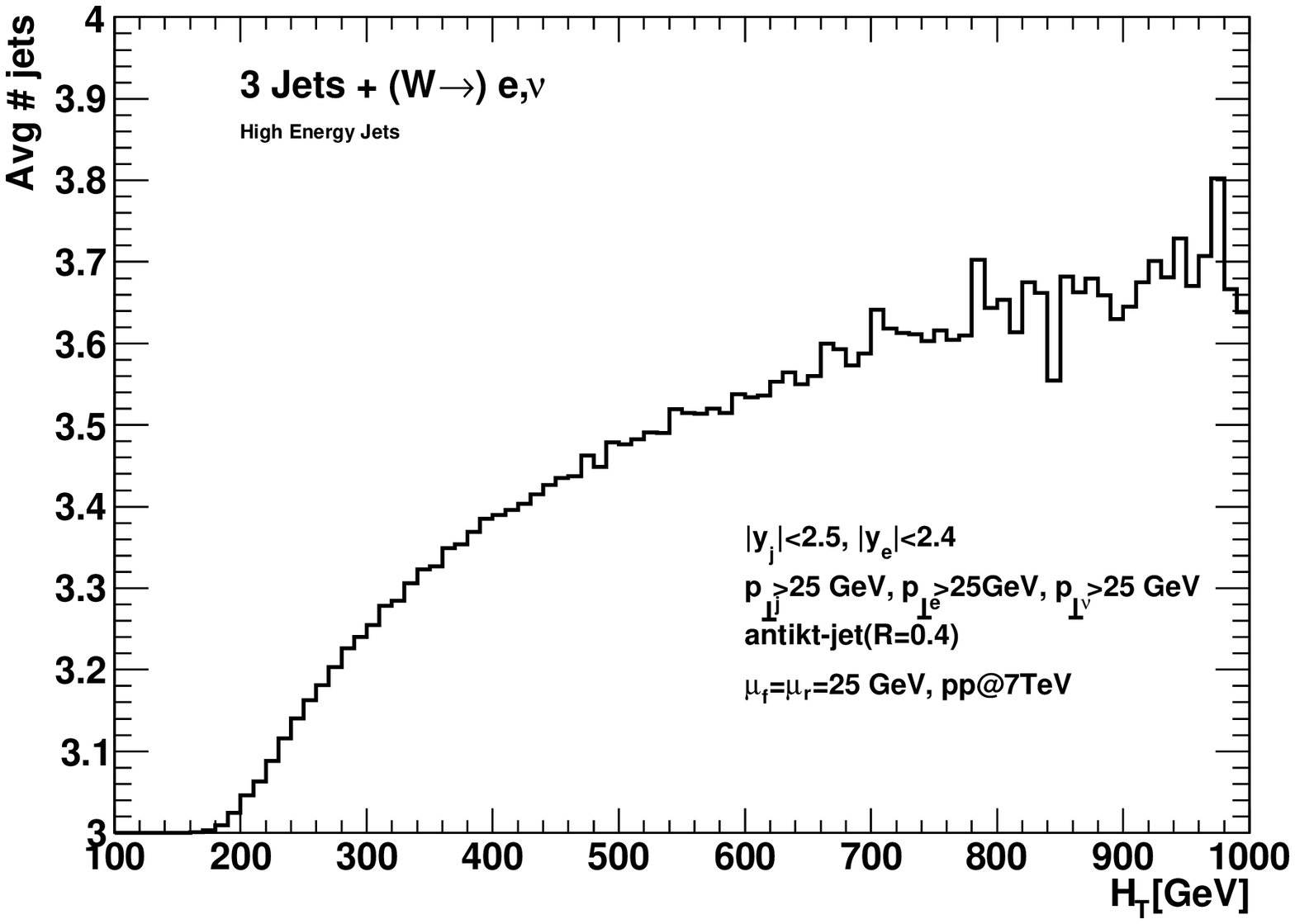}
  \caption{Results obtained with \emph{HEJ} for $W$-production in association
    with at least three jets.}
  \label{fig:w3javgjets}
\end{figure}

In this section we report results obtained for the process of $W$-production
in association with at least three jets, using the cuts of
Ref.\cite{Berger:2009ep} (and mentioned on the plots in
Fig.~\ref{fig:w3javgjets}). Ultimately, a comparative study between the NLO
results and those of \emph{HEJ} is desirable, which would require also
similar choice of renormalisation and factorisation scales etc.

In Fig.~\ref{fig:w3javgjets}(top) we report the average number of jets (with
transverse momentum above 25GeV and rapidities less than 2.5) vs.~the
rapidity difference between the most forward/backward hard jet. Again, we see
a strong correlation, indicative of the increasing phase space for hard
emissions. 
In fact, for $W$+3j-production, $\mathrm{d}\sigma/\mathrm{d}y_{fb}$ peaks at
rapidities of 1-2 units, as illustrated in
Fig.~\ref{fig:w3javgjets}(middle). This would increase to even larger
rapidities, if the rapidity of each jet was allowed to be larger than the
2.5 units allowed in the initial analyses at the LHC.

Finally, in Fig.~\ref{fig:w3javgjets}(bottom) we show the average number of
jets vs.~the scalar sum of transverse momenta $H_T$. There is a strong
correlation between $H_T$ and the average number of hard jets. The tail is
dominated by radiative corrections, so a stable description of the final
state in terms of the number of hard jets is clearly necessary, in order to
reach a stable description of $H_T$ within the SM, and thus assist in the
discrimination between that and any sign of new physics.

\section{CONCLUSIONS}
\label{sec:conclusions}
We have briefly discussed the new all-order framework of \emph{High Energy
  Jets}, and illustrated clear similarities between the jet radiation pattern
in processes of pure jets, and jet production in association with a $W$ or
$H$-boson. A thorough understanding of the increasing relevance of higher-order hard,
perturbative corrections and jet production will clearly be 
important for LHC analyses, and should assist both the analysis of new SM
processes, and the search for new physics.

\subsection*{Acknowledgements}
\label{sec:acknowledgements}
The authors would like to thank the \textsc{BlackHat} collaboration for
ongoing discussions on the phenomenology of $W$+jets. JMS would like to thank
CERN-TH for support at several stages throughout this project. This work is
supported by the EC Marie-Curie Research Training Network ``Tools and
Precision Calculations for Physics Discoveries at Colliders'' under contract
MRTN-CT-2006-035505.

\bibliographystyle{h-elsevier2}
\bibliography{papers}

\end{document}